\documentclass[10pt]{article}
\usepackage{appendix}
\usepackage{epsfig}
\usepackage{capt-of}

\usepackage[activeacute,english]{babel}
\usepackage{latexsym}

\newcommand{\be}{\begin{equation}}
\newcommand{\ee}{\end{equation}}
\newcommand{\bea}{\begin{eqnarray}}
\newcommand{\eea}{\end{eqnarray}}

\newcommand{\gapp}{\mathrel{\raise.3ex\hbox{$>$}\mkern-14mu
              \lower0.6ex\hbox{$\sim$}}}
\newcommand{\lapp}{\mathrel{\raise.3ex\hbox{$<$}\mkern-14mu
              \lower0.6ex\hbox{$\sim$}}}

\newcommand{\lsim}{\mbox{\raisebox{-.9ex}{~$\stackrel{\mbox{$<$}}{\sim}$~}}}

\renewcommand\({\left(}
\renewcommand\){\right)}
\renewcommand\[{\left[}
\renewcommand\]{\right]}

\newcommand\eq[1]{Eq.~(\ref{#1})}

\def\calb{{\cal B}}

\def\calp{{\cal P}}

\def\calt{{\cal T}}

\def\calpz{{\calp_\zeta}}

\def\caltz{{\calt_\zeta}}
\newcommand\bfA{{\mathbf A}}

\newcommand\bfd{{\mathbf d}}

\newcommand\bfk{{\mathbf k}}

\newcommand\bfn{{\mathbf N}}

\newcommand\bfp{{\mathbf p}}
\newcommand\bfq{{\mathbf q}}

\newcommand\bfx{{\mathbf x}}

\newcommand\sub[1]{_{\rm #1}}
\newcommand\su[1]{^{\rm #1}}
\newcommand\mone{^{-1}}

\newcommand\half{^{1/2}}

\newcommand\threehalf{^{3/2}}

\newcommand{\fnl}{f\sub{NL}}

\newcommand{\tnl}{\tau\sub{NL}}

\newcommand\pz{P_\zeta}
\newcommand\bz{B_\zeta}
\newcommand\tz{T_\zeta}

\newcommand\gz{g_\zeta}
\newcommand{\no}{\nonumber}
\newcommand{\dn}{\delta N}

\begin{document}

\title{\begin{flushright}
\normalsize PI/UAN-2009-353FT
\end{flushright}
\vspace{5mm}
{\bf Non-gaussianity from the trispectrum and vector field perturbations}}

\author{
\textbf{C\'esar A. Valenzuela-Toledo$^{1,}$\thanks{e-mail: \texttt{cavalto@ciencias.uis.edu.co}}} and \textbf{Yeinzon Rodr\'{\i}guez$^{1,2,}$\thanks{e-mail: \texttt{yeinzon.rodriguez@uan.edu.co}}}\\ \\
\textit{$^1$Escuela de F\'{\i}sica, Universidad Industrial de Santander,}  \\
\textit{Ciudad Universitaria, Bucaramanga, Colombia} \\
\textit{$^2$Centro de Investigaciones, Universidad Antonio Nari\~no,}\\
\textit{Cra 3 Este \# 47A-15, Bogot\'a D.C., Colombia}\\
}

\maketitle

\begin{abstract}
\noindent We use the $\dn$ formalism to study the trispectrum $\tz$ of the primordial
curvature perturbation $\zeta$ when the latter is generated by vector field perturbations,
considering the tree-level and one-loop contributions. The order of magnitude of the level
of non-gaussianity in the trispectrum, $\tnl$, is calculated in this scenario and related to
the order of magnitude of the level of
non-gaussianity in the bispectrum, $\fnl$, and the level of statistical anisotropy in the
power spectrum, $\gz$. Such consistency relations will put under test this scenario against
future observations. Comparison with the expected observational bound on $\tnl$ from WMAP,
for generic inflationary models, is done.
\end{abstract}

\section{Introduction}
Non-gaussianity in the primordial curvature perturbation $\zeta$ is one of the subjects of more interest in
modern cosmology, because the non-gaussianity parameters $\fnl$ and $\tnl$ together with the spectrum amplitude $A_\zeta$ and
spectral index $n_\zeta$ allow us to discriminate between the different models proposed for the origin of the
large-scale structure (see for example Refs. \cite{lesgourgues,alabidi1,alabidi2,alabidi3}). In most of
these cosmological models it is assumed that the
$n$-point correlators of $\zeta$ are translationally and rotationally invariants. However, since violations of the
translational (rotational) invariance (i.e. violations of the statistical homogeneity (isotropy)) seem to be present in the data
\cite{hou,hoftuft,hansen,dipole1,dipole2,dipole3} (\cite{gawe,ge,app,hl,samal}), many researchers have started to build theoretical models
that include those violations, which
could be due to the presence of vector field perturbations
\cite{armendariz,vc,vc2,RA2,ys,gmv,kksy,gmv2,gvnm,koh,himmetoglu,himmetoglu2,himmetoglu3,himmetoglu4,dklr,dkw,dkw2,go,pitrou},
spinor field perturbations \cite{bohmer,shan,alberghi}, or p-form perturbations \cite{germani,kobayashi,
koivisto,germani2,koivisto2}, due to anisotropic expansion
\cite{kksy,himmetoglu4,bohmer,koivisto,ppu1,gcp,ppu2,watanabe,bamba,dechant} or due to an inhomogeneous background
\cite{armendariz,dklr,carroll}.

Violation of the statistical isotropy is implemented via modifications of the usual
definitions of the statistical descriptors \cite{armendariz,carroll,acw} of the primordial
curvature perturbation $\zeta$. For example, to parametrize the statistical anisotropy under
the assumption of statistical homogeneity, the power spectrum $P'(\bfk)$ must include an
isotropic piece $P(k)$ and an anisotropic piece proportional to the former and exhibiting
explicitly the appearance of a preferred direction: \cite{acw}
\be\label{statan}
P'(\bfk)=P(k)(1+\gz\;(\hat{\bfd}\cdot\hat{\bfk})^2+\ldots) \,.
\ee
In the previous expression $g_\zeta$ is a dimensionless parameter, $\hat{\bfk}$ is the
unitary wave-vector, and $\hat{\bfd}$ is the unitary vector along the preferred direction.
Some recent papers \cite{gawe,ge,app,hl,samal} claim for the presence of statistical
anisotropy in the five-year data from the NASA's WMAP satellite \cite{wmap}. In particular,
if considering just the quadrupolar term of \eq{statan}:
\be
\pz(\bfk) = P_\zeta\su{iso}(k) \( 1 + g_\zeta
(\hat{\bfd}\cdot \hat \bfk)^2 \) \,,
\label{curvquad} \ee
Ref. \cite{gawe} gives
$\gz \simeq 0.290 \pm 0.031$ which rules out statistical isotropy at more than $9 \sigma$.
Nevertheless, the preferred direction lies near the plane of the solar system, which makes
the authors of Ref. \cite{gawe} believe that this effect could be due to an unresolved
systematic error (among other possible systematic errors which have not been demonstrated
either to be the source of this statistical anisotropy nor to be completely uncorrelated
\cite{gawe}).
Even if the result found in Ref. \cite{gawe} turns out to be due to a systematic error, some
forecasted constraints on $\gz$ show that the statistical anisotropy subject is worth
studying \cite{pullen}: $|g_\zeta| \lsim 0.1$ for
the NASA's WMAP satellite \cite{wmap} if there is no detection, and $|g_\zeta| \lsim 0.02$
for the ESA's PLANCK satellite \cite{planck} if there is no detection. 

Recent works show that the particular presence of vector fields in the inflationary dynamics may generate sizeable
levels of non-gaussianity described by $\fnl$ \cite{dkl,bdmr1,vrl} and $\tnl$ \cite{bdmr2}. As shown in Ref.
\cite{vrl}, including vector fields allows us to get consistency relations between the order
of magnitude of the non-gaussianity parameter $\fnl$ and the amount of statistical anisotropy
in the spectrum $\gz$. The above studies may transform the violation of the statistical
isotropy in a decisive tool to discriminate among some of the most usual cosmological models.

In this letter we use the $\dn$ formalism to calculate the tree-level and one-loop contributions to
the trispectrum $\tz$ of $\zeta$ including vector and scalar field perturbations. We then
calculate the order of magnitude of the level of non-gaussianity in $\tz$ including the
one-loop contributions and write down formulas that relate the order of magnitude of $\tnl$
with the amount of statistical anisotropy in the spectrum, $\gz$, and the order of magnitude
of the level of non-gaussianity in the bispectrum, $\fnl$. Finally, comparison with the
expected observational bound from WMAP is done.

\section{Trispectrum from vector field perturbations}		

The $\dn$ formalism \cite{starobinsky2,ss,tanaka,sasaki1,lr1} extended to
include the possible statistical anisotropy in the primordial curvature perturbation $\zeta$
originated from vector field perturbations \cite{dklr}, provides a powerful tool to calculate
$\zeta$ and its statistical descriptors. Assuming an inflationary dynamics
dominated by just one scalar field $\phi$ and one vector field ${\bf A}$, $\zeta$ is expressed
as \cite{dklr}\footnote{This expression
corrects Eq. (3.14) of Ref. \cite{dklr}, and Eq. (3) of Ref. \cite{dkl}, where a factor 2 in
the fourth term of the expansion is missing.}
\be
\zeta(\bfx)\equiv\delta N (\phi(\bfx),A_i(\bfx),t)=N_\phi \delta\phi + N_A^i\delta A_i+\frac{1}{2}N_{\phi\phi}
(\delta\phi)^2+
N_{\phi A}^i\delta\phi\delta A_i+\frac{1}{2}N_{AA}^{ij}\delta A_i \delta A_j \,,\label{deltan}
\ee
where
\be
N_\phi\equiv\frac{\partial N}{\partial \phi}\,,\quad
N_A^{i}\equiv\frac{\partial N}{\partial A_i}\,,\quad
N_{\phi\phi} \equiv\frac{\partial^2 N}{\partial \phi^2}\,,\quad
N_{AA}^{ij}\equiv\frac{\partial^2 N}{\partial A_i\partial A_j}\,,
\quad N_{\phi A}^i\equiv\frac{\partial^2 N}{\partial\phi \partial A_i}\,,
\ee
and $i$ denotes the spatial
indices running from 1 to 3. Now, we define the power spectrum $\pz$ and the
trispectrum $\tz$ for the primordial curvature perturbation, through the Fourier modes of $\zeta$, as:
\bea
\langle\zeta(\bfk_1)\zeta(\bfk_2)\rangle&\equiv&(2\pi)^3\delta(\bfk_1+\bfk_2)P_\zeta(\bfk)
\;\;\equiv \;\;(2\pi)^3\delta(\bfk_1+\bfk_2)\frac{2\pi^2}{k^3}\calpz(\bfk) \,,
\label{spdef}\\
\langle\zeta(\bfk_1)\zeta(\bfk_2)\zeta(\bfk_3)\zeta(\bfk_4)\rangle&\equiv&(2\pi)^3\delta(\bfk_1+\bfk_2+\bfk_3+
\bfk_4)\tz(\bfk_1,\bfk_2,\bfk_3,\bfk_4)\no\\
&\equiv& (2\pi)^3\delta(\bfk_1+\bfk_2+\bfk_3+\bfk_4)\frac{\(2\pi^2\)^3}{k_1^3k_2^3|\bfk_2+\bfk_3|^3}\caltz(\bfk_1,\bfk_2,\bfk_3,\bfk_4) \,.
\label{tsdef}
\eea
As shown in Ref. \cite{dklr}, the tree-level contribution to the spectrum has the form of
\eq{curvquad}. This is simply obtained by using Eqs. (\ref{deltan}) and (\ref{spdef}).
Assuming again only tree-level contributions, the level of non-gaussianity $\fnl$ in the
bispectrum $\bz$ was calculated in Ref. \cite{dkl}. The same calculation was performed in Ref.
\cite{vrl} but this time including also one-loop contributions and considering them to be
dominant over the tree-level terms. In both works, $\pz$ and $\fnl$ were shown to exhibit
anisotropic contributions coming from the vector field perturbation.
In this letter we show that it is possible to obtain an analogous expression for the level of
non-gaussianity $\tnl$ in the trispectrum $\tz$. To do it, we first need to calculate the expressions for $\calpz$
and $\caltz$, defined in Eqs. (\ref{spdef}) and (\ref{tsdef}). Considering contributions up to one-loop order, we
find\footnote{
Eq. (\ref{Pzetal}) corrects a mistake in Eq. (4.12) of Ref. \cite{dklr} where the infinitesimal
volume element $d^3p$ was incorrectly expressed in terms of $dp$.}:
\bea
\calp_\zeta^{\rm tree}(\bfk) &=& N_\phi^2 \calp_{\delta \phi}(k) +
N_A^iN_A^j\calt_{ij}(\bfk)  \nonumber\\
&=&
 N_\phi^2 \calp_{\delta \phi}(k) +
N_A^2 \calp_+(k) + ({\bf N}_A\cdot\hat\bfk )^2 \calp_+(k)  \( r\sub{long} - 1  \)\label{Pzetat}
\,, \eea
\bea
\calp_\zeta^{\rm 1-loop} (\bfk) &=& \int
\frac{d^3p \ k^3}{4\pi|\bfk + \bfp|^3 p^3}
\[\frac{1}{2} N_{\phi \phi}^2  \calp_{\delta \phi} (|\bfk + \bfp|)
\calp_{\delta \phi}(p) +
N_{\phi A}^i N_{\phi A}^j \calp_{\delta \phi} (|\bfk + \bfp|)
\calt_{ij}(\bfp)\right. \nonumber \\&&\left. + \frac{1}{2} N_{AA}^{ij} N_{AA}^{kl}\calt_{ik}(\bfk+\bfp)\calt_{jl}(\bfp) \] \,, \label{Pzetal}
\eea
\bea
\calt_\zeta^{\rm tree} (\bfk_1,\bfk_2,\bfk_3,\bfk_4) &=& N_\phi^2 N_{\phi \phi}^2
[\calp_{\delta \phi} (k_2) \calp_{\delta \phi} (k_4)  \calp_{\delta \phi} (|\bfk_1+\bfk_2|)+ {\rm 11 \ perm.}] \no\\
&+& N_A^i N_A^j N_{AA}^{kl} N_{AA}^{mn}\Big[ \calt_{ik}(\bfk_2)\calt_{jm}(\bfk_4)\calt_{ln}(\bfk_1+\bfk_2) + {\rm 11 \ perm.}\Big] \no\\
 &+& N_\phi^2 N_{A \phi}^i N_{A \phi}^j \Big[\calp_{\delta \phi} (k_2) \calp_{\delta \phi} (k_4)
\calt_{ij}(\bfk_1+\bfk_2) + 11 \ {\rm perm.} \Big] \no\\
&+& N_A^i N_A^j N_{A \phi}^k N_{A \phi}^l \Big[\calt_{ik}(\bfk_2)\calt_{jl}(\bfk_4)\calp_{\delta \phi}(|\bfk_1+\bfk_2|) + 11 \
  {\rm perm.} \Big] \no\\
&+& N_\phi N_{\phi \phi}N_A^i N_{A\phi}^{j}\Big[\calp_{\delta \phi}(k_2)\calt_{ij}(\bfk_4)\calp_{\delta \phi}(|\bfk_1+\bfk_2|) + 23 \
  {\rm perm.} \Big] \no\\
&+& N_\phi N_A^iN_{A\phi}^{j} N_{AA}^{kl} \Big[\calp_{\delta \phi}(k_2)\calt_{ik}(\bfk_4)\calt_{jl}(\bfk_1+\bfk_2) + 23 \ {\rm perm.} \Big]
\,, \label{tst}
\eea
\bea
\calt_{\zeta A}^{\rm 1-loop}(\bfk_1,\bfk_2,\bfk_3,
\bfk_4)&=&N_{AA}^{ij}N_{AA}^{kl}N_{AA}^{mn}N_{AA}^{op}\int\frac{d^3p \ k_1^3k_3^3|\bfk_3+\bfk_4|^3}{4\pi p^3|
\bfk_1-\bfp|^3|\bfk_3+\bfp|^3|\bfk_3+\bfk_4+\bfp|^3}\times\no\\
&&\times\calt_{im}(\bfp)\calt_{jk}(\bfk_1-\bfp)\calt_{np}(\bfk_3+\bfp)\calt_{lo}(\bfk_3+\bfk_4+\bfp)\label{tsl1} \,,
\eea
where
\be\label{def1}
\calt_{ij}(\bfk)\equiv T_{ij}^{\rm even}(\bfk)\calp_+(k)+iT_{ij}^{\rm odd}(\bfk)\calp_-(k)+T\su{long}_{ij}(\bfk)\calp_{\rm long}(k) \,,
\ee
and
\be
T_{ij}\su{even} (\bfk) \equiv \delta_{ij} -  \hat k_i \hat k_j \,,\qquad
T\su{odd}_{ij} (\bfk) \equiv \epsilon_{ijk}\hat k_k \,,\qquad
T\su{long}_{ij} (\bfk) \equiv \hat k_i \hat k_j\,.
\ee
\eq{Pzetat} was written in the form of \eq{curvquad} with $\hat{\bfd}=\hat{\bfn}_A$, $\bfn_A$
being a vector
with magnitude  $N_A\equiv\sqrt{N^i_AN^i_A}$, and $r\sub{long}\equiv\calp\sub{long}/\calp_+$, where $\calp\sub{long}$ is the
power spectrum of the longitudinal component, and $\calp_+$ and $\calp_-$ are the parity conserving and violating
power spectra defined by
\be
\calp_{\pm}\equiv\frac{1}{2}\(\calp_R\pm\calp_L\) \,,
\ee
with $\calp_R$ and $\calp_L$ denoting the power spectra of the transverse components with right-handed and
left-handed polarisations \cite{dklr}. \eq{tsl1} only includes terms coming from vector field perturbations; this
is because the complete expression (including the scalar and the mixed terms) is too large
and in the current paper
we are assuming that the contributions to $T_\zeta$ coming {\it only} from vector fields
dominate over all the other contributions.

\section{Vector field contributions to the statistical descriptors}

When statistical anisotropy is assumed, there is an important restriction from observation: one related to the
amount of statistical anisotropy present in the spectrum, which is given by the parameter $\gz$ in \eq{curvquad}.
Recent studies of the data coming from the WMAP experiment, set an upper bound over $\gz$: $\gz\lsim 0.383$
\cite{gawe}. The latter observational constraint is fully satisfied when we assume that the contributions coming
from vector fields in Eqs. (\ref{Pzetat}) and (\ref{Pzetal}) are smaller than those coming from scalar fields.
That means that the first term in  \eq{Pzetat} dominates over all the other terms, even those coming from one-loop
contributions.

In our study we will assume that the terms coming only from the vector field dominate over those coming from
the mixed terms and from the scalar fields only, except for the case of the tree-level
spectrum, where we will assume that the scalar term is the dominant one\footnote{The power
spectrum $P_\zeta$ must be dominated by the tree-level terms. Otherwise there would be too much
scale dependence in conflict with the current observational limit on $n_\zeta$.}. Of course,
for an actual realisation of this scenario, we need to show that such constraints are fully
satisfied. From the above
assumptions it follows that:
\bea
\calpz^{\rm tree}(\bfk)&=&\calpz_\phi^{\rm tree}(k)+\calpz_ A^{\rm tree}(\bfk)\label{sst1} \,,\\
\calp_\zeta^{\rm 1-loop}(\bfk)&=&\calpz_{A}^{\rm 1-loop}(\bfk)\label{ssl1} \,,\\
\calt_\zeta^{\rm tree} (\bfk_1,\bfk_2,\bfk_3,\bfk_4) &=&\caltz_A^{\rm tree} (\bfk_1,\bfk_2,\bfk_3,\bfk_4)\label{bsst1} \,,\\
\calt_\zeta^{\rm 1-loop}(\bfk_1,\bfk_2,\bfk_3,\bfk_4)&=&\caltz_A^{\rm 1-loop} (\bfk_1,\bfk_2,\bfk_3,\bfk_4)\label{bssl1} \,,
\eea
where the subscripts $\zeta_\phi$ and $\zeta_A$ mean scalar field or vector field
contributions to $\zeta$. The above expressions lead us to two different possibilities that
let us study and probably get a high level of non- gaussianity:
\begin{itemize}
\item Vector field spectrum ($\calp_{\zeta_A}$) and trispectrum ($\calt_{\zeta_A}$) dominated
by the tree-level terms.
\item Vector field spectrum ($\calp_{\zeta_A}$) and trispectrum ($\calt_{\zeta_A}$) dominated
by the one-loop contributions.
\end{itemize}
Other possibilities are not viable because it is impossible to satisfy simultaneously that
the vector field spectrum ($\calp_{\zeta_A}$) is dominated by the tree-level terms and
the trispectrum ($\calt_{\zeta_A}$) is dominated by the one-loop contributions, or the vector field spectrum
($\calp_{\zeta_A}$) is dominated by the one-loop contributions and the trispectrum ($\calt_{\zeta_A}$) is
dominated by the tree-level terms\footnote{See the relevant discussion regarding the vector
field bispectrum $\calb_{\zeta_A}$ in Ref. \cite{vrl}.}. This is perhaps related to the fact that we have taken into account only one
vector field. Such a conclusion may be relaxed if we take into account more than one vector field, as analogously
happens in the scalar multi-field case \cite{cogollo,valenzuela}.

In order to study the above possibilities, we need to estimate the integrals coming from loop
contributions. From Eqs. (\ref{Pzetal}), (\ref{tsl1}), (\ref{ssl1}), and (\ref{bssl1}) the integrals to solve are:
\bea
\calpz^{\rm 1-loop}(\bfk)&=& \frac{1}{2} N_{AA}^{ij} N_{AA}^{kl} \int \frac{d^3p \ k^3}{4\pi p^3|\bfk + \bfp|^3}
\calt_{ik}(\bfk+\bfp)\calt_{jl}(\bfp) \,, \label{intsl}\\
\calt_{\zeta A}^{\rm 1-loop}(\bfk_1,\bfk_2,\bfk_3,
\bfk_4)&=&N_{AA}^{ij}N_{AA}^{kl}N_{AA}^{mn}N_{AA}^{op}\int\frac{d^3p \ k_1^3k_3^3|\bfk_3+\bfk_4|^3}{4\pi p^3|
\bfk_1-\bfp|^3|\bfk_3+\bfp|^3|\bfk_3+\bfk_4+\bfp|^3}\times\no\\
&&\times \calt_{im}(\bfp)\calt_{jk}(\bfk_1-\bfp)\calt_{np}(\bfk_3+\bfp)\calt_{lo}(\bfk_3+\bfk_4+\bfp)\label{tsl} \,. \label{inttsl}
\eea
The above integrals cannot be done analytically, but they can be estimated using the same
technique shown in Appendix
\ref{app}; it is found that the integrals are proportional to $\ln(kL)$
(where $L$ is the box size) if the spectrum is scale invariant.
Following it, we find from Eqs. (\ref{intsl}) and (\ref{inttsl}):
\bea
\calp_{\zeta A}^{\rm 1-loop} (\bfk)&=&\frac{1}{2}N_{AA}^{ij}N_{AA}^{kl}(2\calp_++\calp_{\rm long})\delta_{ik}\calt_{jl}
(\bfk)\ln(kL) \,, \label{sploop}\\
\calt_{\zeta A}^{\rm 1-loop}(\bfk_1,\bfk_2,\bfk_3,\bfk_4)
&=&N_{AA}^{ij}N_{AA}^{kl}N_{AA}^{mn}N_{AA}^{op}\ln(kL)\big(2\calp_++\calp_{\rm long})\delta_{im}\big[\calt_{jk}
(\bfk_1)\calt_{np}(\bfk_3)\calt_{lo}(\bfk_4+\bfk_3)\big] \,. \label{tsploop}
\eea

Observations are available  within
the observable universe and, except for the low multipoles of the CMB, all observations
probe scales $k\gg H_0$. To handle them, one should choose the box size as
$L=H_0\mone$ \cite{leblond}. A smaller choice would throw away some of the data while
a bigger choice would make the spatial averages unobservable.
Low multipoles $\ell$ of the CMB anisotropy explore scales of order $H_0\mone/\ell$ not very much smaller than
$H_0\mone$. To handle them  one has to take $L$ bigger than $H_0\mone$.
For most purposes, one should use a  box, such that $\ln(LH_0)$ is just
a few (i.e.\ not exponentially large) \cite{lythbook,lythbox,klv}.
When comparing the loop contribution with observation one should normally
set $L=H_0\mone$, except for the low CMB multipoles where one should choose $L\gg H_0\mone$ with
$\ln(kL)\sim 1$. With the choice $L=H_0\mone$, $\ln(kL)\sim 5$ for the scales
explored by the CMB multipoles with $\ell \sim 100$, while $\ln(kL)\sim 10$
for the scales explored by galaxy surveys. Since we are interested in giving orders of
magnitude and simple mathematical expressions, we will set $\ln(kL)\sim 1$ without loss of generality.

\section{Calculation of the non-gaussianity parameter $\tnl$}
The non-gaussianity parameter $\tnl$ is defined by \cite{lyth2}:
\be
\tnl=\frac{2\,\caltz(\bfk_1,\bfk_2,\bfk_3,\bfk_4)}{\big[\calpz(\bfk_1)\calpz(\bfk_2)\calpz(\bfk_1+\bfk_4)+{\rm 23\  perm.}\big]}\label{tnl1} \,.
\ee
Remember that the isotropic contribution in the \eq{curvquad} is always dominant compared to the anisotropic one
so that we may write in the above expression only the isotropic part of the spectrum $\calpz^{\rm iso}(k)$:
\be
\tnl=\frac{2\caltz(\bfk_1,\bfk_2,\bfk_3,\bfk_4)}{\big[\calpz^{\rm iso}(k_1)\calpz^{\rm iso}(k_2)\calpz^{\rm iso}(|\bfk_1+\bfk_4|)+{\rm 23\  perm.}\big]} \,.
\label{tnl2}
\ee
Using the above expression, we will estimate the possible amount of non-gaussianity generated by the
anisotropic part of the primordial curvature perturbation, taking into account different possibilities
and assuming that the non-gaussianity is produced solely by vector field perturbations.

\subsection{Vector field spectrum ($\calp_{\zeta_A}$) and trispectrum ($\calt_{\zeta_A}$)
dominated by the tree-level terms}
In this first case, we assume that the trispectrum is dominated by vector field perturbations
and that the higher order terms in the $\dn$ expansion in Eq. (\ref{deltan}) involving the
vector field are sub-dominant against the first-order term:
$N_A^i\delta A_i\gg N_{AA}^{ij}\delta A_i \delta A_j$. The latter implies that both the
spectrum and the trispectrum are dominated by the tree-level terms, i.e.
$\calpz_A^{\rm tree} \gg \calpz_A^{\rm 1-loop}$ and $\caltz_A^{\rm tree} \gg \caltz_A^{\rm 1-loop}$.
Thus, we have from \eq{tnl2}:
\be
\tnl=\frac{2\caltz_A^{\rm tree}(\bfk_1,\bfk_2,\bfk_3,\bfk_4)}{\big[\calpz^{\rm iso}(k_1)\calpz^{\rm iso}(k_2)\calpz^{\rm iso}(|\bfk_1+\bfk_4|)+{\rm 23\  perm.}\big]} \,,
\label{tnl11}
\ee
which, in view of Eqs. (\ref{tst}) and (\ref{bsst1}), looks like:
\be
\tnl\:\simeq\:\frac{2N_A^i N_A^j N_{AA}^{kl} N_{AA}^{mn}\Big[ \calt_{ik}(\bfk_2)\calt_{jm}(\bfk_4)\calt_{ln}(\bfk_1+\bfk_2) + {\rm 11 \ perm.}\Big]}{\big[\calpz^{\rm iso}(k_1)\calpz^{\rm iso}(k_2)\calpz^{\rm iso}(|\bfk_1+\bfk_4|)+{\rm 23\  perm.}\big]} \,.
\label{tnl12}
\ee

We will just consider here the order of magnitude of $\tnl$.  Therefore, we will ignore the
specific ${\bf k}$ dependence of $\calt_{ij}$.  Instead, we will assume that
$\calp_{\rm long}$, $\calp_+$, and $\calp_-$ are all of the same order of
magnitude, which is a good approximation for some specific actions (see for instance Ref.
\cite{dklr}), and take advantage of the fact that the spectrum is almost scale invariant
\cite{wmap5}. Thus, after getting rid of all the ${\bf k}$ dependences, the order of magnitude
of $\tnl$ looks like:
\be
\tnl\:\simeq\:\frac{\calp_A^3 N_A^2 N_{AA}^2}{(\calpz^{\rm iso})^3} \,,
\label{tnl13}
\ee
where $\calp_A=2\calp_++\calp_{\rm long}$. Employing our assumption that $N_A\delta A > N_{AA}\delta A^2$,
and since the root mean squared value for the vector field perturbation $\delta
A$ is $\sqrt{\calp_A}$, the contribution of the vector field to $\zeta$  is given
by $\zeta_A\sim \sqrt{\calpz_A}\sim N_A \sqrt{\calp_A}$. An upper bound for $\tnl$ is therefore
given by:
\be
\tnl\:\lsim\:\frac{\calpz_A^2}{(\calpz^{\rm iso})^3} \,.
\label{tnl14}
\ee
Since the order of magnitude of $\gz$ is $\calpz_A/\calpz^{\rm iso}$, under the assumptions made above
we get:
\be
\tnl\:\lsim\:8\times10^6\(\frac{\gz}{0.1}\)^2 \,,
\label{tnl15}
\ee
where $(\calpz^{\rm iso})\half \simeq 5\times 10 ^{-5}$ \cite{wmap5} 
has been used.
Eq. (\ref{tnl15}) gives an upper bound for the level of non-gaussianity $\tnl$ in terms of
the level of statistical anisotropy in the power spectrum $\gz$ when the former is generated
by the anisotropic contribution to the curvature perturbation. Comparing with the expected
observational limit on $\tnl$ coming from future WMAP data releases, $\tnl \sim2\times 10^4$
\cite{kogo}\footnote{The trispectrum in this scenario might be either of the local, equilateral,
or orthogonal type. We are not interested in this letter on the shape of the non-gaussianity
but on its order of magnitude. Being that the case, comparing with the expected bound on the
{\it local} $\tnl$ \cite{kogo} makes no sensible difference under the assumption that the
expected bounds on the equilateral and orthogonal $\tnl$ are of the same order of magnitude,
as analogously happens in the $\fnl$ case for single-field inflation \cite{smz}.}, we conclude that in this scenario a large level of non-gaussianity in the
trispectrum $T_\zeta$ of $\zeta$ is possible, leaving some room for ruling out this scenario
if the current expected observational limit is overtaken.

As an example of this scenario, we apply the previous results to a specific model, e.g. the vector curvaton
model \cite{vc,vc2,RA2}, where the $N$-derivatives are \cite{dkl}:
\bea
N_{A}&=&\frac{2}{3A}r \,,\label{navc}\\
N_{AA}&=&\frac{2}{A^2}r\label{naavc} \,,
\eea
where $A\equiv|\bfA|$ is the value of vector field just before the vector curvaton field decays
and the parameter $r$ is the ratio between the energy density of the vector curvaton field and
the total energy density of the Universe just
before the vector curvaton decay.

First, we check if the conditions under which the vector field
spectrum and trispectrum are always dominated by the tree-level terms are fully satisfied.
From
Eqs. (\ref{Pzetat}), (\ref{tst}), (\ref{sploop}) and (\ref{tsploop}) our constraint leads to:
\bea
\calp_A N_A^2 &\gg & \calp_A^2 N_{AA}^2 \,, \label{teq} \\
\calp_A^3 N_A^2 N_{AA}^2&\gg &\calp_A^4 N_{AA}^4 \,,
\eea
which mean that the if the vector field spectrum is dominated by the tree-level terms so is
the vector field trispectrum.  An analogous situation happens when the vector field spectrum
is dominated by the one-loop terms:  the vector field trispectrum is also dominated by this
kind of terms. As a result, it is impossible that simultaneously the vector field spectrum
is dominated by the tree-level (one-loop) terms and the vector field trispectrum is dominated
by the one-loop (tree-level) terms.
Following Eq. (\ref{teq}), we get:
\be
\calp_A \ll \(\frac{N_A}{N_{AA}}\)^2 \,,
\ee
which, in view of
$\zeta_A\sim \sqrt{\calpz_A}\sim N_A \sqrt{\calp_A}$ and 
Eqs. (\ref{navc}) and (\ref{naavc}), reduces to:
\be
r \gg 2.25\times 10^{-4}\gz\half \,.\label{rbt}
\ee
This lower bound on the $r$ parameter has to be considered when building a realistic
particle physics model of the vector curvaton scenario.

Second, looking at \eq{tnl13}, we obtain the level of non-gaussianity $\tnl$ for this scenario:
\be
\tnl\simeq \frac{2\times 10^{-2}}{r^2}\(\frac{\gz}{0.1}\)^3 \,.
\ee
This is a consistency relation between $\tnl$, $g_\zeta$, and $r$ which will help when
confronting the specific vector curvaton realisation against observation.  Indeed, a similar
consistency relation between $\fnl$ and $g_\zeta$ was derived for this scenario in Ref.
\cite{vrl}:
\be
\fnl\simeq \frac{4.5\times 10^{-2}}{r}\(\frac{\gz}{0.1}\)^2 \,.
\ee
Thus, in the framework of the vector curvaton scenario, the levels of non-gaussianity $\fnl$
and $\tnl$ are related to each other via the $r$ parameter in this way:
\be
\tnl \simeq \frac{2.1}{r^{1/2}} \fnl^{3/2} \,,
\ee
in contrast to the standard result
\begin{equation}
\tnl = \frac{36}{25} \fnl^2 \,, \label{relbyrnes}
\end{equation}
for the scalar field case (including the scalar curvaton scenario) found in Ref. \cite{bsw}.

\subsection{Vector field spectrum ($\calp_{\zeta_A}$) and trispectrum ($\calt_{\zeta_A}$)
dominated by the one-loop contributions}
From Eqs. (\ref{tsploop}) and (\ref{tnl2}) we get
\be
\tnl\:\simeq\:\frac{N_{AA}^{ij}N_{AA}^{kl}N_{AA}^{mn}N_{AA}^{op}\ln(kL)\big(2\calp_++\calp_{\rm long})\delta_{im}\big[\calt_{jk}
(\bfk_1)\calt_{np}(\bfk_3)\calt_{lo}(|\bfk_4+\bfk_3|)\big]}{\big[\calpz^{\rm iso}(k_1)\calpz^{\rm iso}(k_2)\calpz^{\rm iso}(|\bfk_1+\bfk_4|)+{\rm 23\  perm.}\big]} \,.
\label{tnl21}
\ee

Assuming again that $\calp_{\rm long}$, $\calp_+$, and $\calp_-$ are all of the same order of
magnitude, and that the spectrum is scale invariant, we end up with:
\be
\tnl\:\simeq\:\frac{\calp_A^4 N_{AA}^4}{(\calpz^{\rm iso})^3} \,.
\label{tnl22}
\ee
Performing a similar analysis as done in the previous subsection, but this time taking into
account that the vector field spectrum is dominated by the one-{\rm loop} contribution and
therefore $\zeta_A\sim \sqrt{\calpz_A}\sim
N_{AA} \calp_A$, 
we arrive at:
\be
\tnl\:\sim\:\frac{\calpz_A^2}{(\calpz^{\rm iso})^3}\:\sim\:8\times10^6\(\frac{\gz}{0.1}\)^2.
\label{tnl23}
\ee
The above result gives a relation between the non-gaussianity parameter $\tnl$ and the
level of statistical anisotropy in the power spectrum $\gz$.

Now, we call a similar result that we found for the
non-gaussianity parameter $\fnl$ in Ref. \cite{vrl}, that is:
\be
\fnl\:\sim\:10^3\(\frac{\gz}{0.1}\)\threehalf.
\label{fnlv}
\ee
By combining Eqs. (\ref{tnl23}) and (\ref{fnlv}) we get:
\be
\tnl\:\sim\:8\times10^2\fnl^{4/3},\label{tnlfnl}
\ee
which gives a consistency relation between the non-gaussianity parameters $\fnl$ and $\tnl$ for this
particular scenario. The consistency relations in Eqs. (\ref{tnl23}), (\ref{fnlv}), and
(\ref{tnlfnl}) will put under test this scenario against future observations. In particular,
the consistency relation in Eq. (\ref{tnlfnl}) differs significantly from those obtained when
$\zeta$ is generated only by scalar fields (see e.g. Eq. (\ref{relbyrnes}) and Ref. \cite{bsw}).

Again when we apply our result to the vector curvaton scenario, we get from Eqs. (\ref{Pzetat}), (\ref{tst}),
(\ref{sploop}), (\ref{tsploop}), (\ref{navc}) and (\ref{naavc}) :
\be
r <2.25\times 10^{-4}\gz\half \,, \label{rbl}
\ee
which is an upper bound on the $r$ parameter that must be considered when building a realistic
particle physics model of the vector curvaton scenario.

\section{Conclusions}
We have studied in this letter the order of magnitude of the level of non-gaussianity $\tnl$
in the trispectrum $\tz$ when statistical anisotropy is generated by the presence of one
vector field. We have shown that it is possible to get an upper bound on the order of
magnitude of $\tnl$ if we assume that the tree-level contributions dominate over all higher
order terms in both the vector field spectrum ($\calp_{\zeta_A}$) and the trispectrum
($\calt_{\zeta_A}$); this bound is given in \eq{tnl15}. We also show that it is possible to
get a high level of non-gaussianity $\tnl$, easily exceeding the expected observational
bound from WMAP, if we assume that the one-loop contributions dominate over the tree-level
terms in both the vector field spectrum ($\calp_{\zeta_A}$) and the trispectrum
($\calt_{\zeta_A}$). $\tnl$ is given in this case by \eq{tnl23}, where we may see that there
is a consistency relation between the order of magnitude of $\tnl$ and the amount of
statistical anisotropy in the spectrum $\gz$. Two other consistency relations are given by
Eqs. (\ref{fnlv}) and (\ref{tnlfnl}), this time relating the order of magnitude of the
non-gaussianity parameter $\fnl$ in the bispectrum $\bz$ with the amount of statistical
anisotropy $g_\zeta$ and the order of magnitude of the level of non-gaussianity $\tnl$ in the
trispectrum $\tz$.  Such consistency relations let us fix two of the three parameters by
knowing about the other one, i.e. if the non-gaussianity in the bispectrum (or trispectrum)
is detected and our scenario is appropriate, the amount of statistical anisotropy in the
power spectrum and the order of magnitude of the non-gaussianity parameter $\tnl$ (or $\fnl$)
must have specific values, which are given by Eqs. (\ref{fnlv}) (or (\ref{tnl23})) and
(\ref{tnlfnl}). A similar conclusion is reached if the statistical anisotropy in the
power spectrum is detected before the non-gaussianity in the bispectrum or the trispectrum is.

\appendix
\section{One-loop integral for $P_\zeta$} \label{app}
We sketch in this appendix the mathematical procedure to estimate the integrals in Eqs. (\ref{Pzetal}) and
(\ref{tsl1}). We only work one integral since the other ones are estimated in a similar way.

The one-loop
contribution to the spectrum is:
\bea
\calp_\zeta^{\rm 1-loop} (\bfk) &=& \int
\frac{d^3p \ k^3}{4\pi|\bfk + \bfp|^3 p^3}
\[\frac{1}{2} N_{\phi \phi}^2  \calp_{\delta \phi} (|\bfk + \bfp|)
\calp_{\delta \phi}(p) +
N_{\phi A}^i N_{\phi A}^j \calp_{\delta \phi} (|\bfk + \bfp|)
\calt_{ij}(\bfp)\right. \nonumber \\&&\left. + \frac{1}{2} N_{AA}^{ij} N_{AA}^{kl}\calt_{ik}(\bfk+\bfp)\calt_{jl}(\bfp) \] \,.
\eea
As we can see, the total contribution to $\calp_\zeta^{\rm 1-loop}$ corresponds to three integrals, each one
having two singularities: one in $\bfp=0$ and the other one in $\bfp=-\bfk$. If the fields spectra
are scale invariant, the first integral may be written as:
\be
\calp_\zeta^{\rm 1-loop (a)} (\bfk) = \frac{1}{8\pi}\calp_{\delta \phi}^2 N_{\phi \phi}^2
\int\frac{d^3p \ k^3}{4\pi|\bfk + \bfp|^3 p^3} \,,
\ee
so the actual integral to estimate is:
\be
I=\int_{L^{-1}} \frac{d^3p \ k^3}{|\bfk + \bfp |^3 p^3} \,. \label{dint}
\ee
This integral is logarithmically divergent at the zeros in the denominator,
but there is a cutoff at $k=L^{-1}$. The subscript $L^{-1}$ indicates that the integrand is
set equal to zero
in a sphere of radius $L^{-1}$ around each singularity, and the discussion makes sense only for $L^{-1}\ll k\ll k_{max}$.
If we consider the infrared divergences, that means $\bfp\ll\bfk$, we may write:
\be
I=\int_{L^{-1}}^k\frac{d^3p}{p^3}\sim4\pi\ln(kL).
\ee
To calculate the contribution coming from the other singularity we can make the  substitution $\bfq=\bfk+\bfp$.
After evaluating this latter integral, we find that the contribution is  again $4\pi\ln(kL)$.
The integral in Eq. (\ref{dint}) may be finally estimated by adding the contributions of the two
singularities, that means:
\be
I=\int\frac{d^3p \ k^3}{|\bfk + \bfp |^3 p^3}=8\pi\ln(kL).
\ee
More details to evaluate these integrals may be found in Refs. \cite{lythbox,lyth2,lythaxions}.

The technique to evaluate this kind of integrals when considering vector fields is the same, although
the procedure is algebraically more tedious. Nevertheless, one can finally arrive to the same
conclusion. A more detailed discussion about this issue will be found in a forthcoming
publication \cite{val}.


\renewcommand{\refname}{{\large References}}

\end{document}